\documentclass[prx,aps,onecolumn,amsfonts,showpacs,superscriptaddress,preprint,eqsecnum]
{revtex4-1}
\usepackage{graphicx,amsfonts,times,bm,amsmath,verbatim,color,array}

\usepackage{subfigure}
\usepackage{float}
\usepackage{epsfig}
\usepackage{hyperref}
\hypersetup{
colorlinks=true,final=true,
        linkcolor=blue,
        citecolor=blue,
        filecolor=blue,
        urlcolor=blue,
}

\begin{document}
\title{Berry phase theory of planar Hall effect in Topological Insulators}

\author{S. Nandy}
\affiliation{Department of Physics, Indian Institute of Technology Kharagpur, W.B. 721302, India}
\author{A. Taraphder}
\affiliation{Department of Physics, Indian Institute of Technology Kharagpur, W.B. 721302, India}
\affiliation{Centre for Theoretical Studies, Indian Institute of Technology Kharagpur, W.B. 721302, India}
\author{Sumanta Tewari}
\affiliation{Department of Physics, Indian Institute of Technology Kharagpur, W.B. 721302, India}
\affiliation{Department of Physics and Astronomy, Clemson University, Clemson, SC 29634,U.S.A}


\begin{abstract}
 Negative longitudinal magnetoresistance, in the presence of an external magnetic field parallel to the direction of an applied current, has recently been experimentally verified in Weyl semimetals and topological insulators in the bulk conduction limit. The appearance of negative longitudinal magnetoresistance in topological semimetals is understood as an effect of chiral anomaly, whereas it is not well-defined in topological insulators. Another intriguing phenomenon, planar Hall effect - appearance of a transverse voltage in the plane of applied co-planar electric and magnetic fields not perfectly aligned to each other, a configuration in which the conventional Hall effect vanishes, has recently been suggested to exist in Weyl semimetals.
 In this paper we present a quasi-classical theory of planar Hall effect of a three-dimensional topological insulator in the bulk conduction limit. Starting from Boltzmann transport equations we derive the expressions for planar Hall conductivity and longitudinal magnetoconductivity in topological insulators and show the important roles played by the orbital magnetic moment for the appearance of planar Hall effect. Our theoretical results
 predict specific experimental signatures for topological insulators that can be directly checked in experiments.
\end{abstract}

\maketitle

\section{Introduction}

Three-dimensional (3D) topological insulators (TI) as a new class of quantum matter have recently drawn much attention in condensed matter physics and materials science~\cite{Kane_2010, Zhang_2011}. In TIs, a finite energy gap is present in the bulk, which is crossed by two gapless surface state branches with nontrivial spin textures protected from backscattering by time reversal symmetry. In these systems, owing to strong spin-orbit coupling, electron spins in the surface state branches are aligned perpendicular to their momenta contributing an overall Berry  phase of $\pi$ to the fermion wave functions. In addition to the fundamental scientific interest, TIs evince a wide variety of intriguing transport properties which make them potential candidates for technological applications. For instance, the topological protection of the surface states and the nontrivial spin textures can be of interest for spintronic and quantum computation applications~\cite{Hasan_2010}.

Several transport studies on TIs have revealed various anomalous quantum phenomena associated with the topological surface states, such as the Aharonov-Bohm oscillations in Bi$_{2}$Se$_{3}$ nanoribbons~\cite{Peng_2010}, the weak anti-localization in Bi$_{2}$Se$_{3}$ and Bi$_{2}$Te$_{3}$ thin films~\cite{Chen_2010, He_2011, Hor_2011}, and the two-dimensional SdH oscillations in Bi$_{2}$Te$_{3}$~\cite{Qu_2010}. Very recently, another intriguing phenomenon, negative longitudinal magnetoresistance (LMR) (and conversely, positive longitudinal magnetoconductivity (LMC)) in the presence of parallel electric and magnetic fields, has been discovered from the bulk conduction contribution in 3D topological insulators~\cite{Widemann_2016, Wang_2015, Chan_2012, Wang_2012, He_2013, Taskin_2012}. The observation of this effect in TIs is quite puzzling because the negative LMR in topological semimetals such as Weyl semimetals is widely believed to be due to non-conservation of separate electron numbers of opposite chirality for relativistic massless fermions, an effect known as the chiral or Adler-Bell-Jackiw anomaly~\cite{Goswami:2015,Zhong,Goswami:2013,Adler:1969, Bell:1969, Nielsen:1981, Nielsen:1983, Aji:2012, Zyuzin:2012, Volovik, Wan:2011, Xu:2011, Goswami:2015, Goswami:2013}. Several experimental groups have successfully observed the chiral anomaly induced negative LMR in Dirac and Weyl materials~\cite{He:2014, Liang:2015, CLZhang:2016, QLi:2016, Xiong,Hirsch}. But this picture of chiral anomaly induced negative longitudinal magnetoresistance does not work in topological insulators because chiral anomaly itself is not well defined in these systems.

It was suggested earlier that a positive LMC is not the only effect of chiral anomaly in a topological Weyl semimetal~\cite{Burkov_2017, Nandy_2017}. A second effect of chiral anomaly is the so-called planar Hall effect, i.e., appearance of an in-plane transverse voltage when the co-planar electric and magnetic fields are not perfectly aligned to each other, precisely in a configuration in which the Lorentz force induced conventional Hall effect vanishes. The planar Hall conductivity (PHC) is defined as the transverse conductivity measured along $\hat{y}$, in a direction perpendicular to the applied electric field and current along $\hat{x}$, in the presence of a magnetic field in the $x-y$ plane making an angle $\theta$ with the $x$ axis. This effect is known to occur in ferromagnetic systems~\cite{Ky_1968, Ge_2007} where its origin is non-trivial spin topology. Interestingly, it has also been observed recently in the surface states of a topological insulator where it has been linked to magnetic field induced anisotropic lifting of the protection of the surface states from backscattering~\cite{Taskin_2017}. Our main objective in this work is suggesting the existence of planar Hall effect from the bulk states of 3D topological insulators in the systems exhibiting negative longitudinal magnetoresistance~\cite{Widemann_2016, Wang_2015, Chan_2012, Wang_2012, He_2013, Taskin_2012} using semi-classical Boltzmann transport theory incorporating topological Berry phase effects. In a complete theory we also derive the associated expressions for longitudinal magnetoconductivity in topological insulators in the bulk conduction limit~\cite{Dai_2017} and predict specific magnitude and direction dependence of PHC and LMC on applied field that can be tested in experiments.

With the motivation for this work described above, in this paper we have chosen Bi$_{2}$Se$_{3}$ as a reference 3D strong topological insulator and study the PHC and LMC expected from its bulk states. This material has been clearly identified as a 3D strong topological insulator with a bulk band gap of 0.3 eV, with a single spin-helical Dirac cone on each surface, which has been confirmed in angle-resolved photoemission spectroscopy measurements ~\cite{Xia_2009}.
Our work on planar Hall effect in this systems, together with positive longitudinal magnetoconductance, completes the quasi-classical description of Berry curvature induced anomalous magneto-transport phenomena in three dimensional topological insulators.

The rest of the paper is organized as follows. In Sec.~\ref{Model Hamiltonian}, we introduce the effective Hamiltonian for the bulk states of a 3D strong topological insulator Bi$_{2}$Se$_{3}$. In Sec.~\ref{Boltzmann}, we derive the analytical expressions of LMC and PHC using semiclassical Boltzmann transport equations. In sec.~\ref{LMC and PHC}, we show our numerical results on LMC and PHC establishing the anomalous features in the transport properties. We also make comparison of our results with existing experimental data. Finally in Sec.~\ref{summary}, we discuss the experimental aspects of the phenomena observed in our study and end with a brief conclusion.

\section{Model Hamiltonian}
\label{Model Hamiltonian}
The low-energy long-wavelength properties of a 3D topological insulator in the presence of time reversal and space inversion symmetries can be described by the effective $\mathbf{k} \cdot \mathbf{p}$ Hamiltonian,
\begin{eqnarray}
{\cal H}_{0}(\mathbf{k})=\begin{pmatrix} \begin{array}{cccc} \epsilon_{\mathbf{k}}+M_{\mathbf{k}} & 0 & iV_{n}k_{z} & -iV_{\parallel}\mathbf{k_{-}}\\0 & \epsilon_{\mathbf{k}}+M_{\mathbf{k}} & iV_{\parallel}\mathbf{k_{+}} & iV_{n}k_{z}\\-iV_{n}k_{z} & -iV_{\parallel}\mathbf{k_{-}} & \epsilon_{\mathbf{k}}-M_{\mathbf{k}} & 0\\iV_{\parallel}\mathbf{k_{+}} & -iV_{n}k_{z} & 0 & \epsilon_{\mathbf{k}}-M_{\mathbf{k}} \end{array} \end{pmatrix} \nonumber \\
\label{H_bulk}
\end{eqnarray}
\begin{figure}[htb]
\begin{center}
\epsfig{file=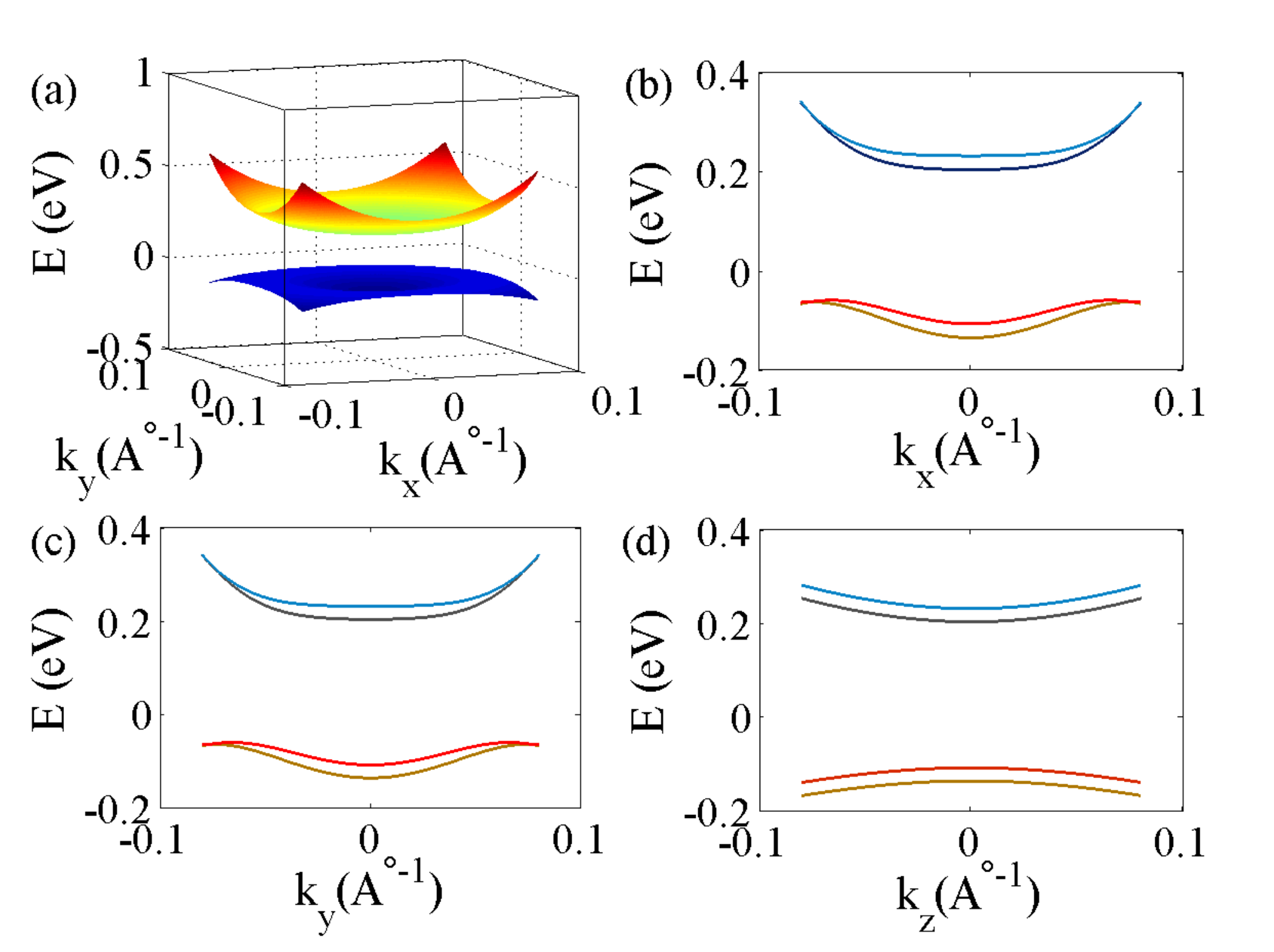,trim=0.0in 0.05in 0.0in 0.05in,clip=true, width=120mm}\vspace{0em}
\caption{(Color online) (a) 3D band dispersion of the four bands ($k_{z}$ is suppressed) of 3D topological insulator (Bi$_{2}$Se$_{3}$) near $\Gamma$ point obtained by diagonalizing Hamiltonian described in Eq.~(\ref{H_bulk}). The doubly degenerate valence bands are separated from the doubly degenerate conduction bands by an energy equal to 0.3 eV. (b)-(d) depict the 2D band dispersions of the same four bands as mentioned above along $k_{x}$, $k_{y}$, and $k_{z}$ axis respectively in the presence of a Zeeman field of strength 5 T applied along the $x$ direction. The Land\'{e} g-factors used here are $g_{pv}=g_{pc}=20$~\cite{Konczykowski_2016}.}
\label{disperse}
\end{center}
\end{figure}
where $\epsilon_{\mathbf{k}}=\epsilon_{0}+\epsilon_{\parallel}\mathbf{k}_{\parallel}^{2}+\epsilon_{z}k_{z}^{2}$, $M_{\mathbf{k}}=M_{0}+M_{\parallel}\mathbf{k}_{\parallel}^{2}+M_{z}k_{z}^{2}$, and $k_{\parallel}^{2}=k_{x}^{2}+k_{y}^{2}$ with $C_{i}$, $M_{i}$, and $V_{i}$ as model parameters. In the present work we have taken $C_{0}=0.048$ eV, $C_{\parallel}=13.9$ eV-${\mbox{\normalfont\AA}^{2}}$, $C_{z}=1.409$ eV-${\mbox{\normalfont\AA}^{2}}$, $M_{0}=-0.169$ eV, $M_{\parallel}=29.36$ eV-${\mbox{\normalfont\AA}^{2}}$, $M_{z}=3.351$ eV-${\mbox{\normalfont\AA}^{2}}$, $V_{n}=1.853$ eV-${\mbox{\normalfont\AA}}$, and $V_{\parallel}=2.512$ eV-${\mbox{\normalfont\AA}}$ to represent the Bi$_{2}$Se$_{3}$ topological insulator as suggested by the ab-initio bandstructure calculations~\cite{Zhang_2009, Krasovskii_2016}. The Hamiltonian described by the Eq.~\ref{H_bulk} includes uniaxial anisotropy along the $z$-direction and k-dependent mass terms. $M_{0}M_{\parallel} < 0$ and $M_{0}M_{z} < 0$ implies that this model belongs to a 3D strong topological insulator~\cite{shen_2012}. The 3D dispersion of the four bulk bands near the $\Gamma$ point, two degenerate conduction bands and valence bands each, is depicted in Fig.~\ref{disperse}(a). To compute the longitudinal conductivity and planar Hall conductivity in the bulk conduction limit we have assumed that the Fermi level crosses the conduction bands.

In addition to the band energy, the Berry curvature $\mathbf{\Omega(k)}$ of the Bloch bands is required for a complete description of the electron dynamics in topological systems. The general form of the Berry curvature can be obtained via symmetry analysis. Under time reversal symmetry, the Berry curvature satisfies $\mathbf{\Omega(-k)}=-\mathbf{\Omega(k)}$. On the other hand, if the system also has spatial inversion symmetry, then $\mathbf{\Omega(-k)}=\mathbf{\Omega(k)}$. Therefore, for 3D topological insualtors like Bi$_{2}$Se$_{3}$ with simultaneous time reversal and spatial inversion symmetries, the Berry curvature vanishes identically throughout the Brillouin zone~\cite{Xiao_2010}. However, in the presence of a magnetic field applied in the $x-y$ plane, the Berry curvature attains non-zero values because of broken time reversal symmetry.

To study the Berry curvature induced magneto-transport phenomena in the presence of an in-plane magnetic field we add to Eq.~(\ref{H_bulk}) the Zeeman magnetic term given by,
\begin{eqnarray}
{\cal H}_{z}=\frac{\mu_{B}}{2}\begin{pmatrix} \begin{array}{cccc} 0 & g_{pv}B_{-} & 0 & 0\\g_{pv}B_{+} & 0 & 0 & 0\\0 & 0 & g_{pc}B_{-} & 0\\0 & 0 & 0 & g_{pc}B_{+} \end{array} \end{pmatrix}
\label{H_field}
\end{eqnarray}
where $\mu_{B}$ is Bohr magneton and $g_{pv}$, $g_{pc}$ are the Land\'{e} g factors for valence and conduction bands in the $x-y$ plane respectively. The 2D dispersions of the valence and conduction bands of the topological insulator along $k_{x}$, $k_{y}$, and $k_{z}$ in the presence of an in-plane magnetic field of strength 5 T applied along $x$ axis are shown in Fig.~\ref{disperse}(b)-(d) respectively. It is clear from the figure that the dispersions of the four bands along $k_{x}$ and $k_{y}$ are identical. The Zeeman splitting of conduction bands is maximum ($\sim$ 28 meV) at the $\Gamma$ point.

\section{Boltzmann Equation Approach For Planar Hall effect}
\label{Boltzmann}
In this section, we derive the semiclassical formulae for the planar Hall conductivity and longitudinal electrical conductivity (LEC) in the low field regime starting from the quasi-classical Boltzmann transport equations. For completeness we include the effects of the orbital magnetic moment $\mathbf{m}$, which is the angular momentum of the semi-classical wave packet and also of geometrical origin, modifying the expressions for LEC and PHC significantly. The complete theory produces the magnetic field and direction dependence of longitudinal magnetoconductivity and planar Hall conductivity in topological insulators that can be verified in experiments.

In the presence of an electric field ($\mathbf{E}$) and temperature gradiant ($\mathbf{\nabla T}$), the charge current ($\mathbf{J}$) and the thermal current ($\mathbf{Q}$) flowing through the system can be described by the linear response equations,
\begin{eqnarray}
\begin{pmatrix} \begin{array}{c} \mathbf{J} \\  \mathbf{Q}\end{array} \end{pmatrix} = \begin{pmatrix} \begin{array} {cc} \hat{\sigma} &  \hat{\alpha} \\ \hat{\bar{\alpha}} & \hat{l} \end{array} \end{pmatrix} \begin{pmatrix} \begin{array} {c} \mathbf{E} \\ -\mathbf{\nabla T}\end{array}  \end{pmatrix}
\label{e1}
\end{eqnarray}
where $\hat{\sigma}$, $\hat{\alpha}$, and $\hat{l}$ are different conductivity tensors. The tensors $\hat{\bar{\alpha}}$ and $\hat{\alpha}$ are related to each other by Onsager's relation $\hat{\bar{\alpha}}=T\hat{\alpha}$. In the linear response theory, we can write J and Q as
\begin{eqnarray}
J_{a}=\sigma_{ab}E_{b}+\alpha_{ab}(-\nabla_{b} T)
\label{e2a}
\end{eqnarray}
\begin{align}
Q_{a}=T\alpha_{ab}E_{b}+l_{ab}(-\nabla_{b} T)
\label{e2b}
\end{align}
The phenomenological Boltzmann transport equation in the presence of impurity scattering can be written as~\cite{John_2001}
\begin{equation}
\left(\frac{\partial}{\partial t}+\mathbf{\dot{r}}\cdot\mathbf{\nabla_{r}}+\mathbf{\dot{k}}\cdot\mathbf{\nabla_{k}}\right)f_{\mathbf{k},\mathbf{r},t}=I_{coll} \{f_{\mathbf{k},\mathbf{r},t}\}
\label{e3}
\end{equation}
where on the right side $I_{coll}$ is the collision integral which incorporates electron correlations
and impurity scattering effects and $f_{\mathbf{k},\mathbf{r},t}$ is the electron distribution function.
Using relaxation time approximation, the collision integral takes the form $I_{coll}\{f_{\mathbf{k}}\}=\frac{f_{eq}-f_{k}}{\tau(\mathbf{k})}$, where $\tau (\mathbf{k})$ is the relaxation time and $f_{eq}$ is the equilibrium Fermi-Dirac distribution function in the absence of any external fields. In this paper we have ignored momentum dependence of $\tau$ and assume the parameter to be a constant in the semiclassical limit for simplifying the calculation~\cite{Burkov_2014}. Dropping the $\mathbf{r}$ dependence of $f_{\mathbf{k},\mathbf{r},t}$, valid for spatially uniform fields, and assuming steady state the Boltzmann equation described by Eq.(~\ref{e3}) takes the following form
\begin{equation}
(\mathbf{\dot{r}}\cdot\mathbf{\nabla_{r}}+\mathbf{\dot{k}}\cdot\mathbf{\nabla_{k}})f_{k}=\frac{f_{eq}-f_{\mathbf{k}}}{\tau(\mathbf{k})}
\label{e4}
\end{equation}

In the presence of electric field and magnetic field, transport properties get substantially modified due to presence of non-trivial Berry curvature which acts as a fictitious magnetic field in the momentum space~\cite{Xiao_2010}. The Berry curvature of the $n^{th}$ band for a Bloch Hamiltonian $H(\mathbf{k})$, defined as the Berry phase per unit area in the $\mathbf{k}$ space, is given by
\begin{equation}
\Omega_{\mu\nu}^{n}(\mathbf{k})=2i\sum \frac{<n|\frac{\partial H}{\partial k^{\mu}}|n^{\prime}><n^{\prime}|\frac{\partial H}{\partial k^{\nu}}|n>}{(\epsilon_{n}-\epsilon_{n^{\prime}})^{2}}
\label{e5}
\end{equation}
The wave packet of a Bloch electron also carries an orbital magnetic moment in addition to its spin moment due to the self rotation around its center of mass. The orbital magnetic moment associated with $n^{th}$ Bloch band can be defined as~\cite{Niu_1999}
\begin{equation}
m_{\mu\nu}^{n}(\mathbf{k})=-\frac{ie}{\hbar}\sum \frac{<n|\frac{\partial H}{\partial k^{\mu}}|n^{\prime}><n^{\prime}|\frac{\partial H}{\partial k^{\nu}}|n>}{\epsilon_{n}-\epsilon_{n^{\prime}}}
\label{e6}
\end{equation}
It is clear from the above equation that the orbital magnetic moment does not depend on the actual shape and size of the wave packet but only depends on the Bloch functions. The orbital moment has exactly the same symmetry properties as the Berry curvature, namely, $m(-\mathbf{k})=-m(\mathbf{k})$ and $m(-\mathbf{k})=m(\mathbf{k})$ under time reversal and inversion symmetries, respectively. Therefore $m(\mathbf{k})$ vanishes in the simultaneous presence of both these symmetries. In the present case, the orbital moment is non-zero because of broken time reversal symmetry due to the in-plane magnetic field.

 As the orbital moment couples to the magnetic field ($\mathbf{B}$) through a Zeeman-like term $-\mathbf{m(k)} \cdot \mathbf{B}$, the unperturbed band energy $\epsilon_{0,\mathbf{k}}$ is modified as $\epsilon_{\mathbf{k}}=\epsilon_{0,\mathbf{k}}-\mathbf{m(k)} \cdot \mathbf{B}$. In the presence of $m(\mathbf{k})$ the group velocity of Bloch electrons is also modified as $\mathbf{\tilde{v}_{k}}=\mathbf{v_{k}}-\frac{1}{\hbar}\nabla(\mathbf{m}\cdot \mathbf{B})$. Incorporating the effects due to Berry curvature and orbital magnetic moment, the semi-classical equation of motion for an electron takes the following form~\cite{Niu_1999}
\begin{align}
\mathbf{\dot{r}}=\frac{1}{\hbar}\nabla\epsilon_{\mathbf{k}}+\mathbf{\dot{k}}\times\mathbf{\Omega_{k}}
\label{e49}
\end{align}
\begin{align}
\hbar\mathbf{\dot{k}}=e\mathbf{E}+e\mathbf{\dot{r}} \times \mathbf{B}
\label{e50}
\end{align}
where the second term of the Eq.~(\ref{e49}) implies the anomalous velocity originating from the non-trivial Berry curvature. The coupled equations for $\mathbf{\dot{r}}$ and $\mathbf{\dot{k}}$ described in Eq.~(\ref{e49}) and Eq.~(\ref{e50}) can be solved together to obtain~\cite{Moore_2016}
\begin{align}
\mathbf{\dot{r}}=D(\mathbf{B,\Omega_{k}})[\mathbf{\tilde{v}_{k}}+\frac{e}{\hbar}(\mathbf{E}\times\mathbf{\Omega_{k}})+\frac{e}{\hbar}(\mathbf{\tilde{v}_{k}}\cdot\mathbf{\Omega_{k}})\mathbf{B}]
\label{7a}
\end{align}
\begin{align}
\hbar\mathbf{\dot{k}}=D(\mathbf{B,\Omega_{k}})[e\mathbf{E}+\frac{e}{\hbar}(\mathbf{\tilde{v}_{k}} \times \mathbf{B})+\frac{e^{2}}{\hbar}(\mathbf{E}\cdot\mathbf{B})\mathbf{\Omega_{k}}]
\label{7b}
\end{align}
Here the prefactor $D(\mathbf{B,\Omega_{k}})=(1+\frac{e}{\hbar}(\mathbf{B}.\mathbf{\Omega_{k}}))^{-1}$, modifying the invariant phase space volume according to $dkdx \rightarrow D(\mathbf{B,\Omega_{k}})dkdx$, gives rise to a noncommutative mechanical model, because the Poisson brackets of co-ordinates
is nonzero~\cite{Duval_2006}. For ease of notation we will simply denote $D(\mathbf{B,\Omega_{k}})$ by $D$ for rest of the paper. 

The second term of the Eq.~(\ref{7a}) gives rise to the anomalous transport induced by the Berry curvature.
The third term in the same equation gives rise to chiral magnetic effect modified by the orbital magnetic moment. The chiral magnetic effect, an interesting signature of transport phenomena in Weyl semimetals, appears in equilibrium i.e. $\mathbf{E}=0$~\cite{Son_2012, Yin_2012, Chen_2013, Kenji_2008}. This term implies an electric current $\propto \mathbf{B}$ to flow along the direction of the magnetic field in Weyl semimetals without any eletric field in the presence of finite chiral chemical potential ($\mu_{+}-\mu_{-}$) where $\mu_{+}$ and $\mu_{-}$ are the chemical potentials of two Weyl nodes~\cite{Kim:2014}. There has been some controversy regarding the existence of the equilibrium chiral magnetic effect in condensed matter systems because the effect described above violates the Maxwell's equations~\cite{Franz_2013, Basar_2014, Landsteiner_2014, Chen_2013}. It has been discussed that in the dc limit i.e. when frequency is set to zero first, the system is in equilibrium and the chiral magnetic effect vanishes~\cite{Chen_2013}.
The second term in Eq.~(\ref{7b}) implies the usual Lorentz force modified by $m(\mathbf{k})$ whereas the last term proportional to $\mathbf{E} \cdot \mathbf{B}$ in Eq.~(\ref{7b}) is the semi-classical manifestation of the topological effect known as chiral anomaly in the context of topological semimetals. The chiral anomaly in topological Weyl semimetals implies the non-conservation of a chiral current i.e. violation of the separate number conservation law of Weyl Fermions of a given chirality in the presence of parallel electric and magnetic fields. It is important to note that chiral anomaly is a purely quantum mechanical effect, and while the third term in Eq.~(\ref{7b}) has been interpreted in the literature as the semi-classical manifestation of chiral anomaly in topological semimetals, the term itself may be non-zero in the presence of non-trivial Berry curvature even in systems that do not support chiral anomaly in the quantum limit.

To calculate planar Hall conductivity, we apply an electric field ($\mathbf{E}$) along the $x-$axis and a magnetic field ($\mathbf{B}$) in the $x-y$ plane at a finite angle $\theta$ from the $x-$axis, i.e. $\mathbf{B}=B\cos\theta\hat{x}+B\sin\theta\hat{y}$, $\mathbf{E}=E\hat{x}$. Here, $\theta$ is the angle between $\mathbf{E}$ and $\mathbf{B}$. After substituting $\mathbf{\dot{r}}$ and $\mathbf{\dot{k}}$ into the Boltzmann equation Eq.~(\ref{e4}), it then takes the form,
\begin{align}
&D[(\frac{eE\tilde{v}_{x}}{\hbar}+\frac{e^{2}}{\hbar}BE\cos\theta(\mathbf{\tilde{v}_{k}}.\mathbf{\Omega_{k}}))\frac{\partial \tilde{f}_{eq}}{\partial \tilde{\epsilon}}+\frac{eB}{\hbar^{2}}(-\tilde{v}_{z}\sin\theta\frac{\partial}{\partial k_{x}} \nonumber \\
&(\tilde{v}_{x}\sin\theta-\tilde{v}_{y}\cos\theta)\frac{\partial}{\partial k_{z}}+\tilde{v}_{z}\cos\theta\frac{\partial}{\partial k_{y}})\tilde{f}_{\mathbf{k}}] =\frac{\tilde{f}_{eq}-\tilde{f_{\mathbf{k}}}}{\tau}
\label{e8}
\end{align}
where $\tilde{f}_{eq}$ is equilibrium Fermi-Dirac distribution with the energy dispersion $\epsilon_{\mathbf{k}}=\epsilon_{0,\mathbf{k}}-\mathbf{m} \cdot \mathbf{B}$ modified due to the orbital magnetic moment. 

We now attempt to solve the above equation by assuming the following ansatz for the electron distribution function $\tilde{f}_{\mathbf{k}}$
\begin{eqnarray}
\tilde{f}_{\mathbf{k}}=\tilde{f}_{eq}-eDE\tau({\tilde{v}_{x}}+\frac{eB\cos \theta}{\hbar}(\mathbf{\tilde{v}_{k}}\cdot\mathbf{\Omega_{k}})-\mathbf{\tilde{v}}\cdot\mathbf{\Gamma})\left(\frac{\partial \tilde{f}_{eq}}{\partial\tilde{\epsilon}}\right) \nonumber \\
\label{e9}
\end{eqnarray}
where $\mathbf{\Gamma}$ is the correction factor due to the finite magnetic field $\mathbf{B}$. Inserting $\tilde{f}_{\mathbf{k}}$ into the Eq.~(\ref{e8}), we have
\begin{eqnarray}
& \frac{eB}{\hbar^{2}}\left(-\tilde{v}_{z}\sin\theta\frac{\partial}{\partial k_{x}}+\tilde{v}_{z}\cos\theta\frac{\partial}{\partial k_{y}}+(\tilde{v}_{x}\sin\theta-\tilde{v}_{y}\cos\theta)\frac{\partial}{\partial k_{z}}\right) \nonumber \\
&(eED\tau(\tilde{v}_{x}+\frac{eB\cos\theta}{\hbar}(\mathbf{\tilde{v}_{k}}\cdot\mathbf{\Omega_{k}}))+
\mathbf{\tilde{v}}\cdot\mathbf{\Gamma})=\frac{\mathbf{\tilde{v}}\cdot\mathbf{\Gamma}}{D\tau}
\label{e10}
\end{eqnarray}
The correction factors $\Gamma_{x}$, $\Gamma_{y}$, and $\Gamma_{z}$, calculated by using the fact that the above equation is valid for all values of the velocity, are too small in magnitude compared to the other terms in $\tilde{f}_{\mathbf{k}}$. Neglecting these correction factors we can now rewrite the Boltzmann distribution function $\tilde{f}_{\mathbf{k}}$ as,
\begin{align}
\tilde{f}_{\mathbf{k}}=\tilde{f}_{eq}-eDE\tau({\tilde{v}_{x}}+\frac{eB\cos \theta}{\hbar}(\mathbf{\tilde{v}_{k}}\cdot\mathbf{\Omega_{k}}))(\frac{\partial \tilde{f}_{eq}}{\partial\tilde{\epsilon}})
\label{e11}
\end{align}

Now, in the absence of any thermal gradient, we write the charge density ($\rho$) and current density ($J$) as,~\cite{Moore_2016}
\begin{align}
\rho=e\int [d\mathbf{k}] D^{-1}\tilde{f}_{\mathbf{k}}
\label{e51}
\end{align}
\begin{align}
\mathbf{J}=e\int [d\mathbf{k}](D^{-1}\mathbf{\dot{r}}+\nabla_{r} \times m_{\mathbf{k}})\tilde{f}_{\mathbf{k}}
\label{e52}
\end{align}
where $[d\mathbf{k}]=\frac{d^{3}\mathbf{k}}{(2\pi)^{3}}$ and the factor D arises from a field-induced change in the volume of the phase space. The second term of Eq.~(\ref{e52}) is a contribution of magnetization current. As we are working with spatially uniform fields, in the present work the expression for the current density takes the following form,
\begin{align}
\mathbf{J}=e\int [d\mathbf{k}][\mathbf{\tilde{v}_{k}}+\frac{e}{\hbar}(\mathbf{E}\times\mathbf{\Omega_{k}})+\frac{e}{\hbar}(\mathbf{\tilde{v}_{k}}\cdot\mathbf{\Omega_{k}})\mathbf{B}]\tilde{f}_{\mathbf{k}}
\label{e88}
\end{align}
Plugging $\tilde{f}_{k}$ into the above equation and comparing it with Eq.~(\ref{e2a}), we now arrive at the semiclassical formula for the longitudinal electrical conductivity up to the second order in B including the effects due to Berry curvature and orbital magnetic moment,
\begin{align}
\sigma_{xx}&=e^{2}\int\frac{d^{3}k}{(2\pi)^{3}}\tau [D({\tilde{v}_{x}}+\frac{eB\cos \theta}{\hbar}(\mathbf{\tilde{v}_{k}}\cdot\mathbf{\Omega_{k}}))^{2}]\left(-\frac{\partial \tilde{f}_{eq}}{\partial \tilde{\epsilon}}\right)
\label{e12}
\end{align}
In the above equation the anomalous velocity factor $\frac{eB\cos \theta}{\hbar}(\mathbf{\tilde{v}_{k}}\cdot\mathbf{\Omega_{k}})$ appears due to the topological term ($\mathbf{E} \cdot \mathbf{B}$) and the orbital magnetic moment. This term is the origin of finite B-dependent longitudinal electrical conductivity which is independent of $\mathbf{B}$ for a regular Fermi liquid. To separate the various B-dependent contributions to $\sigma_{xx}$, we write Eq.~(\ref{e12}) in a slightly different form,
\begin{align}
\sigma_{xx}=\sigma_{xx}^{(0)}+\sigma_{xx}^{(1)}+\sigma_{xx}^{(2)}
\label{e13}
\end{align}
where $\sigma_{xx}^{(0)}$, $\sigma_{xx}^{(1)}$, and $\sigma_{xx}^{(2)}$ represent the zeroth order, linear, and quadratic order dependence on B.  Now $\sigma_{xx}^{(0)}$, the B-independent conductivity, can be written as,
\begin{eqnarray}
\sigma_{xx}^{(0)}=-e^{2}\int[dk]\tau v_{x}^{2}\left(\frac{\partial \tilde{f}_{eq}}{\partial \tilde{\epsilon}}\right)
\label{e14}
\end{eqnarray}
The second term $\sigma_{xx}^{(1)}$, describing the linear dependence on B, takes the form,
\begin{align}
\sigma_{xx}^{(1)}&=\frac{2e^{2}}{\hbar}\int[dk]B\tau v_{x}[\nabla_{x}M-e(\mathbf{v_{k}} \cdot \mathbf{\Omega_{k}})\cos \theta]\left(\frac{\partial \tilde{f}_{eq}}{\partial \tilde{\epsilon}}\right) \nonumber \\
&+\frac{e^{3}}{\hbar}\int[dk]\tau v_{x}^{2}(\mathbf{B} \cdot \mathbf{\Omega_{k}})\left(\frac{\partial \tilde{f}_{eq}}{\partial \tilde{\epsilon}}\right)
\label{e15}
\end{align}
where $M=m_{x}\cos \theta + m_{y} \sin \theta$. The third term $\sigma_{xx}^{(2)}$, quadratic in B, is given by
\begin{align}
\sigma_{xx}^{(2)}&=\frac{2e^{4}}{\hbar^{2}}\int[dk]B\tau v_{x} \cos \theta(\mathbf{v_{k}} \cdot \mathbf{\Omega_{k}})(\mathbf{B} \cdot \mathbf{\Omega_{k}})\left(\frac{\partial \tilde{f}_{eq}}{\partial \tilde{\epsilon}}\right) \nonumber \\
&+\frac{2e^{3}}{\hbar^{2}}\int[dk]B^{2}\tau \cos \theta v_{x}(\mathbf{\nabla_{k}}M \cdot \mathbf{\Omega_{k}})\left(\frac{\partial \tilde{f}_{eq}}{\partial \tilde{\epsilon}}\right) \nonumber \\
&+\frac{2e^{3}}{\hbar^{2}}\int[dk]B^{2}\tau \cos \theta \nabla_{x}M (\mathbf{v_{k}}\cdot    \mathbf{\Omega_{k}})\left(\frac{\partial \tilde{f}_{eq}}{\partial \tilde{\epsilon}}\right) \nonumber \\
&-\frac{e^{2}}{\hbar^{2}}\int[dk]B^{2}\tau v_{x}^{2}(\nabla_{x}M)^{2}\left(\frac{\partial \tilde{f}_{eq}}{\partial \tilde{\epsilon}}\right) \nonumber \\
&-\frac{e^{4}}{\hbar^{2}}\int[dk]B^{2}\tau \cos^{2} \theta(\mathbf{v_{k}} \cdot \mathbf{\Omega_{k}})^{2} \left(\frac{\partial \tilde{f}_{eq}}{\partial \tilde{\epsilon}}\right) \nonumber \\
&-\frac{2e^{3}}{\hbar^{2}}\int[dk]B\tau v_{x}\nabla_{x}M (\mathbf{B} \cdot \mathbf{\Omega_{k}})\left(\frac{\partial \tilde{f}_{eq}}{\partial \tilde{\epsilon}}\right) \nonumber \\
\label{e16}
\end{align}
When $\theta=0$, we recover the formula for LEC for parallel $\mathbf{E}$ and $\mathbf{B}$ fields as derived in earlier works~\cite{Son:2013, Kim:2014, Lundgren:2014, Sharma:2016}.

Now we will derive the expression of PHC. Inserting $\tilde{f}_{k}$ in Eq.~(\ref{e88}) and comparing it with Eq.~(\ref{e2a}), we write the following expression for the electrical Hall conductivity up to the second order in B,
\begin{align}
\sigma_{yx}&=e^{2}\int\frac{d^{3}k}{(2\pi)^{3}}D\tau\left(-\frac{\partial \tilde{f}_{eq}}{\partial \tilde{\epsilon}}\right) [(\tilde{v}_{y}+\frac{eB\sin \theta}{\hbar}(\mathbf{\tilde{v}_{k}}\cdot\mathbf{\Omega_{k}})) \nonumber \\
&(\tilde{v}_{x}+\frac{eB\cos \theta}{\hbar}(\mathbf{\tilde{v}_{k}}\cdot\mathbf{\Omega_{k}}))]-\frac{e^{2}}{\hbar}\int\frac{d^{3}k}{(2\pi)^{3}}\Omega_{z}\tilde{f}_{eq}
\label{e17}
\end{align}
We extract the various B-dependent contributions to $\sigma_{yx}$ by writing the above equation as,
\begin{align}
\sigma_{yx}=\sigma_{yx}^{(0)}+\sigma_{yx}^{(1)}+\sigma_{yx}^{(2)}
\label{e18}
\end{align}
where
\begin{align}
\sigma_{yx}^{(0)}=e^{2}\int[dk]\tau v_{x}v_{y}\left(-\frac{\partial \tilde{f}_{eq}}{\partial \tilde{\epsilon}}\right)-\frac{e^{2}}{\hbar}\int[dk]
\Omega_{z}\tilde{f}_{eq}
\label{e19}
\end{align}
gives the B independent Hall contribution with modification due to orbital magnetic moment. The second term of Eq.~(\ref{e19}) gives the anomalous Hall conductivity which arises in the absence of an external magnetic field. This term actually vanishes in the inversion symmetry breaking Weyl semimetals whereas it gives a finite contribution in the present case. The linear and quadratic $B$-dependent contributions to the planar Hall conductivity are given by,
\begin{align}
\sigma_{yx}^{(1)}&=\frac{e^{3}}{\hbar}\int[dk]\tau v_{y} v_{x}(\mathbf{B} \cdot \mathbf{\Omega_{k}}) \left(-\frac{\partial \tilde{f}_{eq}}{\partial \tilde{\epsilon}}\right) \nonumber \\
&-\frac{e^{3}}{\hbar}\int[dk]B \tau (v_{y} \cos \theta + v_{x} \sin \theta) (\mathbf{v_{k}} \cdot \mathbf{\Omega_{k}}) \left(\frac{\partial \tilde{f}_{eq}}{\partial \tilde{\epsilon}}\right)\nonumber \\
&+\frac{e^{2}}{\hbar}\int[dk]B \tau (v_{y}\nabla_{x}M+v_{x}\nabla_{y}M) \left(\frac{\partial \tilde{f}_{eq}}{\partial \tilde{\epsilon}}\right)
\label{e20}
\end{align}
and
\begin{align}
\sigma_{yx}^{(2)}&=-\frac{e^{2}}{\hbar}\int[dk]B^{2}\tau \nabla_{x}M \nabla_{y}M \left(\frac{\partial \tilde{f}_{eq}}{\partial \tilde{\epsilon}}\right)+\frac{e^{3}}{\hbar^{2}}\int[dk] B\tau \nonumber \\
&[B(v_{y} \cos \theta + v_{x} \sin \theta) (\mathbf{\nabla_{k}}M \cdot \mathbf{\Omega_{k}})-(v_{y} \nabla_{x}M + v_{x} \nabla_{y}M) \nonumber \\ &(\mathbf{B} \cdot \mathbf{\Omega_{k}})+ B(\cos \theta \nabla_{y}M+ \sin \theta \nabla_{x}M) (\mathbf{v_{k}} \cdot \mathbf{\Omega_{k}})](\frac{\partial \tilde{f}_{eq}}{\partial \tilde{\epsilon}}) \nonumber \\
&-\frac{e^{4}}{\hbar^{2}}\int[dk]B\tau(\mathbf{v_{k}} \cdot \mathbf{\Omega_{k}})[(v_{x} \sin \theta+v_{y} \cos \theta) (\mathbf{B} \cdot \mathbf{\Omega_{k}})  \nonumber \\
&-B\sin \theta \cos \theta (\mathbf{v_{k}} \cdot
\mathbf{\Omega_{k}})]\left(\frac{\partial \tilde{f}_{eq}}{\partial \tilde{\epsilon}}\right) \nonumber \\
\label{e21}
\end{align}
respectively. As in the present paper we are primarily interested in planar Hall conductivity, we will not consider the anomalous Hall contribution to the total conductivity. Neglecting the anomalous Hall term in Eq.~(\ref{e17}) we then arrive at our final expression for the planar Hall conductivity,
\begin{align}
\sigma_{yx}^{\text{ph}}&=e^{2}\int\frac{d^{3}k}{(2\pi)^{3}}D\tau\left(-\frac{\partial \tilde{f}_{eq}}{\partial \tilde{\epsilon}}\right) [(\tilde{v}_{y}+\frac{eB\sin \theta}{\hbar}(\mathbf{\tilde{v}_{k}}\cdot\mathbf{\Omega_{k}})) \nonumber \\
&(\tilde{v}_{x}+\frac{eB\cos \theta}{\hbar}(\mathbf{\tilde{v}_{k}}\cdot\mathbf{\Omega_{k}}))]
\label{e80}
\end{align}
When $m=0$ we recover the formula for planar Hall conductivity as discussed in earlier work~\cite{Nandy_2017}.
\section{Longitudinal magnetoconductivity and planar Hall conductivity in topological insulators}
\label{LMC and PHC}
In this section we show the B-dependence and angular dependence of longitudinal magnetoconductivity and planar Hall conductivity computed using Eq.~(\ref{e12}) and Eq.~(\ref{e80}), respectively. Negative longitudinal magnetoresistance has recently been observed in several topological insulators in the presence of bulk conduction~\cite{Widemann_2016, Wang_2015, Chan_2012, Wang_2012, He_2013, Taskin_2012}.
Although the planar Hall conductivity has recently been observed from the surface states of a 3D topological insulator~\cite{Taskin_2017}, it is not observed from bulk states till date. In the present work we consider only bulk  states and neglect the contribution to conductivity from the surface states. 

\begin{figure}[htb]
\begin{center}
\epsfig{file=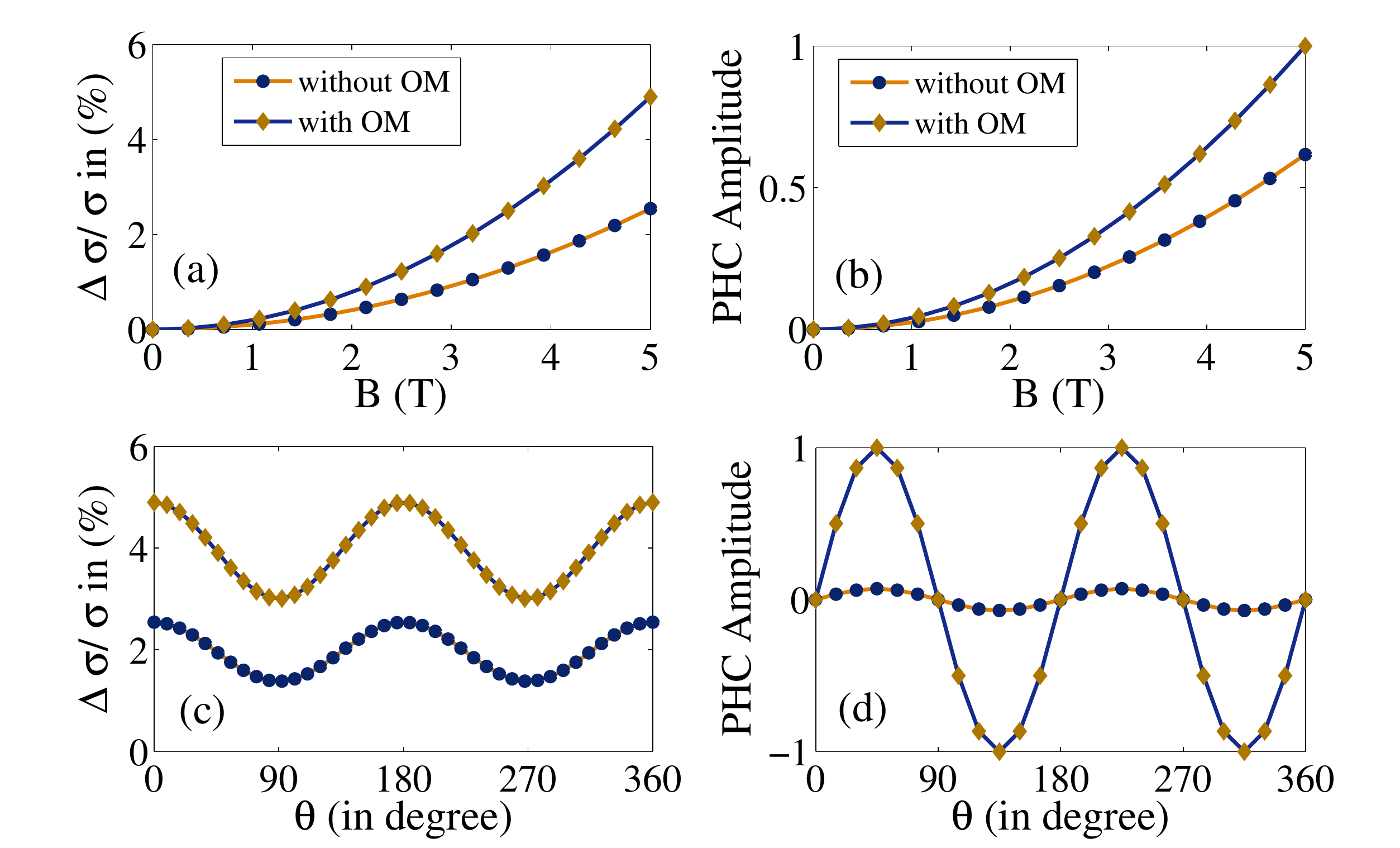,trim=0.0in 0.05in 0.0in 0.05in,clip=true, width=140mm}\vspace{0em}
\caption{(Color online) (a) depicts the LMC in the presence as well as absence of orbital magnetic moment ($m$) as a function of in-plane magnetic field at temperature $T=24$ K and $\theta=0^{\circ}$. (b) shows the dependence of amplitude of PHC (normalized by the maximum value of PHC in the presence of $m$) as a function of in-plane magnetic field at $\theta=\pi/4$. The other parameters are same as above. (c)-(d) show the angular dependence of LMC and PHC (normalized) in the presence and absence of $m$ for $B=5$ T. Here we have normalized the y axis of (d) by the value of PHC at $\theta=\pi/4$ in the presence of $m.$ Curves in yellow indicate the presence of $m$ whereas blue lines are for $m=0$. In all cases  we consider the Fermi level situated at 27 meV from the bottom of the lowest conduction band.}
\label{ang_dep}
\end{center}
\end{figure}
In Fig.~\ref{ang_dep}(a) we have plotted the LMC as a function of the applied magnetic field at $T=24$ K in the presence and absence of $m$ where we have defined LMC as
\begin{eqnarray}
\frac{\Delta \sigma}{\sigma}=\frac{\sigma_{xx}(B)-\sigma_{xx}(B=0)}{\sigma_{xx}(B=0)}
\label{e22}
\end{eqnarray}
The LMC increases monotonically with the magnetic field in both cases and follows the B$^{2}$-dependence.  The orbital moment, first-order correction to the classical equations of motion, increases the Zeeman splitting between two conduction bands and enhances the LMC significantly. Therefore it is essential to take into account the effect of $m$ in computing magnetoconductivity for topological insulators. The first-order contribution described by Eq.~(\ref{e15}) is finite in the present case of topological insulators, indicating the remarkable fact that the non trivial Berry curvature and orbital magnetic moment can produce an anisotropy in the magnetoconductivity even without the chiral anomaly effect. The LMC also follows $\cos^{2} \theta$ dependence at $B=5$ T in both cases (presence and absence of $m$) as depicted in Fig.~\ref{ang_dep}(c), leading to the anisotropic magnetoresistance (AMR).
\begin{figure}[htp]
\centering
\begin{tabular}{cc}
\includegraphics[width=90mm]{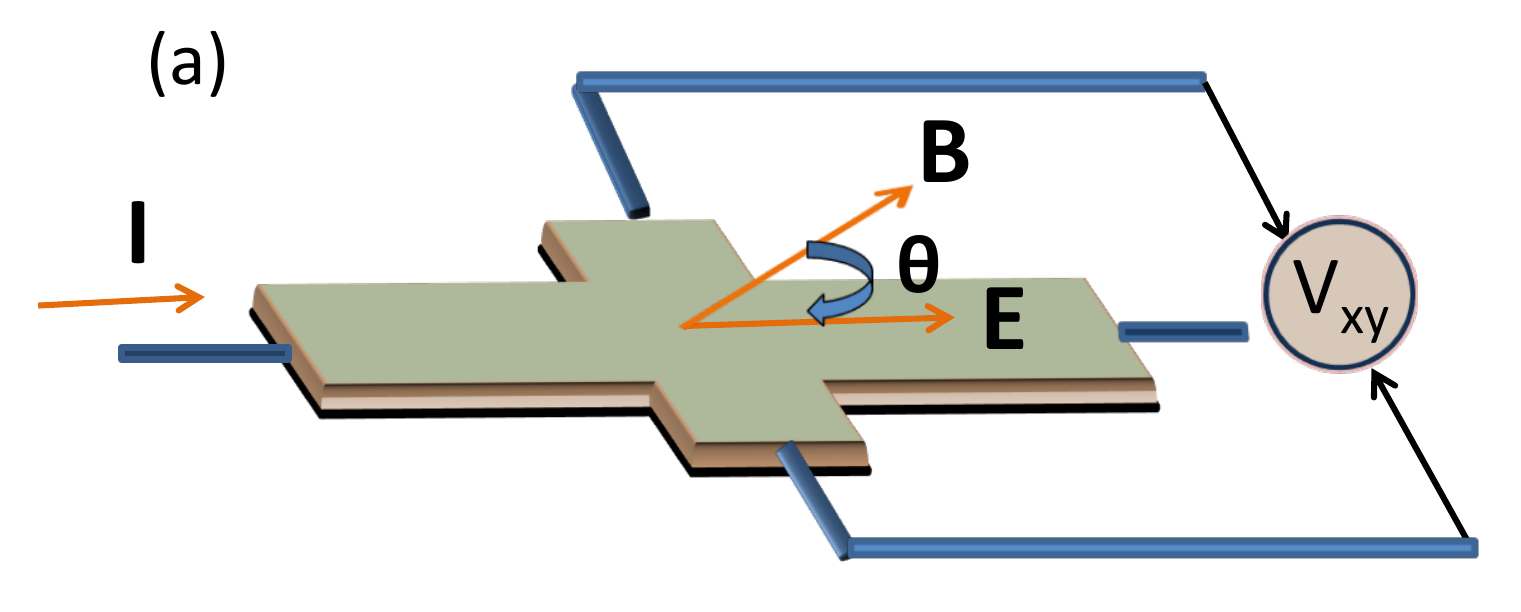}
\includegraphics[width=60mm]{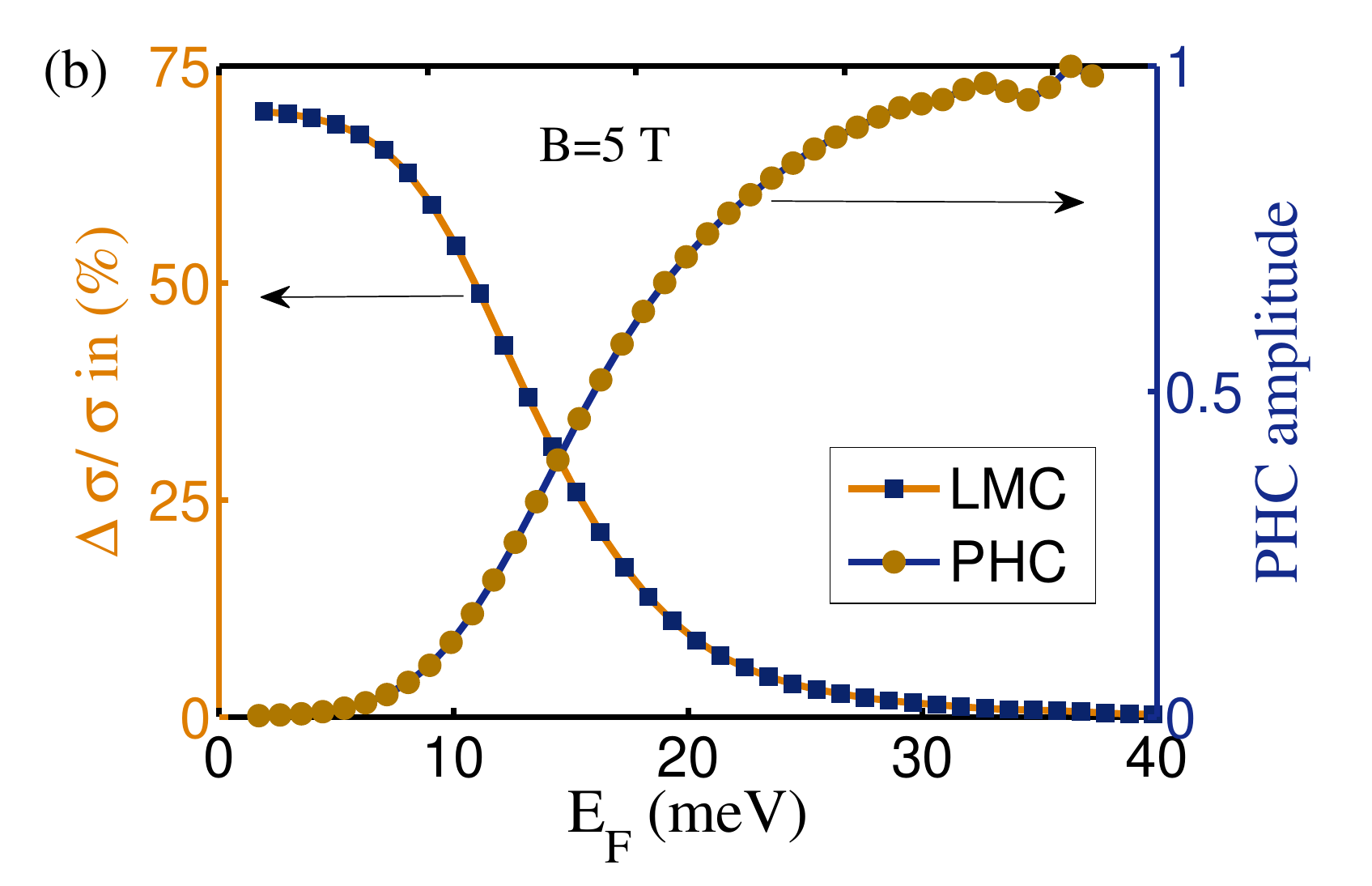}
\end{tabular}
\label{gate}
\caption{(Color online) (a) Illustration for the planar Hall effect measurement geometry. The electric field ($\mathbf{E}$) is applied along the $x-$axis and a magnetic field ($\mathbf{B}$) in the $x-y$ plane makes a finite angle $\theta$ with the $\mathbf{E}$. The planar Hall effect is measured as an in-plane voltage ($V_{xy}$) transverse to the direction of current in the $x-y$ plane. (b) shows the LMC ($\theta=0$) and amplitude of planar Hall conductivity ($\theta=\frac{\pi}{4}$) as a function of Fermi energy for $B=5$ T and $T=24$ K for the bulk states of 3D topological insulators. It is important to note that the magnitudes of LMC and PHC behave very differently as the band filling approaches the bottom of the conduction bands. The Fermi energy is measured from the bottom of the lowest conduction band. The amplitude of PHC has been normalized by its maximum value at $E_{F}=45$ meV.}
\end{figure}
In Fig.~\ref{ang_dep}(b) and Fig.~\ref{ang_dep}(d) we have shown the amplitude and angular dependence of planar Hall conductivity in topological insulators for the bulk states conduction in Bi${_2}$Se${_3}$. The amplitude of the PHC is finite at all field directions except at $\theta=0$ and $\theta=\pi/2$ and follows a quadratic dependence on $B$ which is similar to what has been observed in experiments on PHC on the surface states of Bi${_2}$Se${_3}$. The amplitude is enhanced significantly due to the presence of $m$ leading to the fact that orbital moment plays a very important role in PHC. The planar Hall conductivity $\sigma_{xy}^{\text{ph}}$ does not satisfy the familiar anti-symmetry relation ($\sigma_{xy} = - \sigma_{yx}$) in the spatial indices and this property can be used to identify PHC in experiments. Within the regime of applicability of quasi-classical formalism, we have found that the PHC follows ~$\cos \theta \sin \theta$ dependence for $B=5$ T as depicted in Fig.~\ref{ang_dep}(d). This is also similar what has been observed in experiments on PHC due to surface states of Bi${_2}$Se${_3}$.

Fig. 3(a) depicts schematic diagram of planar Hall effect measurement geometry. In the presence of orbital magnetic moment, the LMC and PHC as a function of Fermi energy ($E_{F}$) for $B=5$ T and $T=24$ K are shown in Fig. 3(b). In experiment, the Fermi level can be tuned by gate voltage. It is clear from the figure that the LMC is enhanced as the Fermi level approaches the band bottom whereas the amplitude of PHC decreases.

\section{Conclusions}
\label{summary}
In this work we present a quasiclassical theory of planar Hall conductivity due to bulk conduction in 3D strong topological insulators (Bi$_{2}$Se$_{3}$) using the phenomenological Boltzmann transport theory. In the presence of in-plane electric and magnetic fields not perfectly aligned with each other, we find a non-zero planar Hall response which is very different in nature from the usual Lorentz force mediated Hall response and even the Berry phase mediated anomalous Hall response, both of which are antisymmetric in spatial indices. We have derived an analytical expression for planar Hall conductivity taking into account the orbital magnetic moment along with the non trivial Berry curvature. 
Our results imply that both LMC and PHC appear in topological insulators due to non-trivial Berry curvature and orbital magnetic moment of the conduction bands. Our numerical results predict experimental observations of PHC together with LMC from the bulk states of 3D strong topological insulators which can be tested in experiments.

\section{Acknowledgement}
 The authors (SN and AT) acknowledge the computing facility from DST-Fund for S and T infrastructure (phase-II) Project installed in the Department of Physics, IIT Kharagpur, India. SN  acknowledges  MHRD, India for support. ST acknowledges support from ARO Grant No: (W911NF-16-1-0182).


\begin{thebibliography}{10}

\bibitem{Kane_2010} M. Z. Hasan and C. L. Kane, Rev. Mod. Phys. \textbf{82}, 3045 (2010).

\bibitem{Zhang_2011} X. L. Qi and S. C. Zhang, Rev. Mod. Phys. \textbf{83}, 1057 (2011).

\bibitem{Hasan_2010} Hasan, M. Z.; Kane, C. L. Colloquium, Rev. Mod. Phys. \textbf{82}, 3045–3067 (2010).

\bibitem{Peng_2010} H. L. Peng, K. J. Lai, D. S. Kong, S. Meister, Y. L. Chen, X. L. Qi, S. C. Zhang, Z. X. Shen, and Y. Cui, Nat. Mater. \textbf{9}, 225 (2010).

\bibitem{Chen_2010} J. Chen, H. J. Qin, F. Yang, J. Liu, T. Guan, F. M. Qu, G. H. Zhang, J. R.
Shi, X. C. Xie, C. L. Yang, K. H. Wu, Y. Q. Li, and L. Lu, Phys. Rev. Lett. \textbf{105}, 176602 (2010).

\bibitem{He_2011} H. T. He, G. Wang, T. Zhang, I. K. Sou, G. K. L. Wong, J. N. Wang, H. Z. Lu, S. Q. Shen, and F. C. Zhang, Phys. Rev. Lett. \textbf{106}, 166805 (2011).

\bibitem{Hor_2011} J. G. Checkelsky, Y. S. Hor, R. J. Cava, and N. P. Ong, Phys. Rev. Lett. \textbf{106}, 196801 (2011).

\bibitem{Qu_2010} D. X. Qu, Y. S. Hor, J. Xiong, R. J. Cava, and N. P. Ong, Science \textbf{329}, 821 (2010).

\bibitem{Widemann_2016} S. Wiedmann, A. Jost, B. Fauque, J. van Dijk, M. J. Meijer, T. Khouri, S. Pezzini, S. Grauer, S. Schreyeck, C. Brune, H. Buhmann, L. W. Molenkamp, and N. E. Hussey, Phys. Rev. B \textbf{94}, 081302 (2016).

\bibitem{Chan_2012} Jian Wang, Handong Li, Cuizu Chang, Ke He, Joon Sue Lee, Haizhou Lu, Yi Sun, Xucun Ma, Nitin Samarth, Shunqing Shen, Qikun Xue, Maohai Xie, and Moses H.W. Chan, Nano Research \textbf{5} 739-746 (2012).

\bibitem{Wang_2015} Li-Xian Wang, Yuan Yan, Liang Zhang, Zhi-Min Liao, Han-Chun Wu, and Da-Peng Yu, Nanoscale \textbf{7}, 16687-16694 (2015).

\bibitem{Wang_2012} Jian Wang, Handong Li, Cuizu Chang, Ke He, JoonSue Lee, Haizhou Lu, Yi Sun, Xucun Ma, Nitin Samarth, Shunqing Shen, Qikun Xue, Maohai Xie, and Moses H.W. Chan, Nano Research \textbf{5}, 739-746 (2012).

\bibitem{He_2013} H. T. He, H. C. Liu, B. K. Li, X. Guo, Z. J. Xu, M. H. Xie, and J. N. Wang, Appl. Phys. Lett. \textbf{103}, 031606 (2013).

\bibitem{Taskin_2012} A. A. Taskin, Satoshi Sasaki, Kouji Segawa, and Yoichi Ando, Phys. Rev. Lett., \textbf{109}, 066803 (2012).

\bibitem{Volovik} G. E. Volovik, Universe in a helium droplet, (Oxford University Press, 2003).

\bibitem{Wan:2011} X. Wan, A. M. Turner, A. Vishwanath, and S. Y. Savrasov, Phys. Rev. B \textbf{83}, 205101 (2011).

\bibitem{Xu:2011} G. Xu, H. Weng, Z. Wang, X. Dai, and Z. Fang, Phys. Rev. Lett. \textbf{107}, 186806 (2011).

\bibitem{Nielsen:1981} H. B.  Nielsen and M. Ninomiya, Phys. Lett. B \textbf{105} 219 (1981).

\bibitem{Nielsen:1983} H. B. Nielsen and M. Ninomiya, Phys. Lett. B \textbf{130}, 389 (1983).

\bibitem{Goswami:2015} P. Goswami, G. Sharma, S. Tewari, Phys. Rev. B \textbf{92}, 161110 (2015).

\bibitem{Zhong} S. Zhong, J. Orenstein, J. E. Moore, Phys. Rev. Lett. \textbf{115}, 117403 (2015).

\bibitem{Goswami:2013} P. Goswami and S. Tewari, Phys. Rev. B \textbf{88}, 245107 (2013).

\bibitem{Bell:1969} J. S. Bell and R. A. Jackiw, Nuovo Cimento A \textbf{60}, 47 (1969).

\bibitem{Aji:2012} V. Aji, Phys. Rev. B \textbf{85} 241101 (2012).

\bibitem{Adler:1969} S. Adler, Phys. Rev. \textbf{177}, 2426 (1969).

\bibitem{Zyuzin:2012} A. A. Zyuzin, S. Wu, and A. A. Burkov, Phys. Rev. B \textbf{85}, 165110 (2012).

\bibitem{He:2014} L. P. He, X. C. Hong, J. K. Dong, J. Pan, Z. Zhang, J. Zhang, and S. Y. Li, Phys. Rev. Lett. \textbf{113}, 246402 (2014).

\bibitem{Liang:2015} T. Liang, Q. Gibson, M. N. Ali, M. Liu, R. J. Cava, N. P. Ong,  Nat Mater \textbf{14}, 280 (2015).

\bibitem{CLZhang:2016} C.-L. Zhang, S.-Y. Xu, I. Belopolski, Z. Yuan, Z. Lin, B. Tong, G. Bian, N. Alidoust, C.-C. Lee, S.-M. Huang, T.-R. Chang, G. Chang, C.-H. Hsu, H.-T. Jeng, M. Neupane, D. S. Sanchez, H. Zheng, J. Wang, H. Lin, C. Zhang, H.-Z. Lu, S.-Q. Shen, T. Neupert, M. Z. Hasan, and S. Jia, Nat. Commun. \textbf{7}, 10735 (2016).

\bibitem{QLi:2016} Q. Li, D. E. Kharzeev, C. Zhang, Y. Huang, I. Pletikosic, A. V. Fedorov, R. D. Zhong, J. A. Schneeloch, G. D. Gu, and T. Valla, Nat. Phys. 12, 550 (2016).

\bibitem{Xiong} J. Xiong, S. K. Kushwaha, T. Liang, J. W. Krizan, M. Hirschberger, W. Wang, R. J. Cava, and N. P. Ong, Science \textbf{350}, 413 (2015).

\bibitem{Hirsch} M. Hirschberger, S. Kushwaha, Z. Wang, Q. Gibson, S. Liang, C. A. Belvin, B. A. Bernevig, R. J. Cava, and N. P. Ong, Nat Mater \textbf{15}, 1161 (2016).

\bibitem{Burkov_2017} A. A. Burkov, arXiv:1704.05467 (2017).

\bibitem{Nandy_2017} S. Nandy, G. Sharma, A. Taraphder, and Sumanta Tewari, arXiv:1705.09308 (2017).

\bibitem{Ky_1968} Vu Dinh Ky, Phys. stat. sol. \textbf{26}, 565 (1968).

\bibitem{Ge_2007} Z. Ge, W. L. Lim, S. Shen, Y. Y. Zhou, X. Liu, J. K. Furdyna, and M. Dobrowolska, Phys. Rev. B \textbf{75}, 014407 (2007).

\bibitem{Taskin_2017} A. A. Taskin, Henry F. Legg, Fan Yang,1 Satoshi Sasaki, Yasushi Kanai, Kazuhiko Matsumoto, Achim Rosch, and Yoichi Ando, arXiv:1703.03406 (2017).

\bibitem{Dai_2017} Xin Dai, Z. Z. Du, and Hai-Zhou Lu, arXiv:1705.02724 (2017).

\bibitem{Xia_2009} Y. Xia, D. Qian, D. Hsieh, L. Wary, A. Pal, H. Lin, A. Bansil, D. Grauer, Y. S. Hor, R. J. Cava, and M. Z. Hasan, Nat. Phys. \textbf{5}, 398 (2009).

\bibitem{Zhang_2009} Haijun Zhang, Chao-Xing Liu, Xiao-Liang Qi, Xi Dai, Zhong Fang, and Shou-Cheng Zhang, Nature Phys. \textbf{5}, 438-442 (2009).

\bibitem{Krasovskii_2016} I. A. Nechaev and E. E. Krasovskii, Phys. Rev. B \textbf{94}, 201410 (2016).

\bibitem{shen_2012} Shun-Qing Shen, Topological Insulators (Springer-Verlag, Berlin Heidelberg, 2012).

\bibitem{Konczykowski_2016} A. Wolos, S. Szyszko, A. Drabinska, M. Kaminska, S. G. Strzelecka, A. Hruban, A. Materna, M. Piersa, J. Borysiuk, K. Sobczak, and M. Konczykowski, Phys. Rev. B \textbf{93}, 155114 (2016).

\bibitem{Zhang_2010} Chao-Xing Liu, Xiao-Liang Qi, HaiJun Zhang, Xi Dai, Zhong Fang, and Shou-Cheng Zhang Phys. Rev. B \textbf{82}, 045122 (2010).

\bibitem{Xiao_2010} D. Xiao, M. C. Chang, and Q. Niu, Rev. Mod. Phys. \textbf{82}, 1959 (2010).

\bibitem{John_2001} John. M. Ziman, Electrons and phonons: the theory of transport
phenomena in solids. Oxford, UK: Clarendon Press, (2001).

\bibitem{Burkov_2014} A. A. Burkov, Phys. Rev. Lett., \textbf{113}, 247203 (2014).

\bibitem{Niu_1999} Ganesh Sundaram and Qian Niu, Phys. Rev. B \textbf{59}, 14915-14925 (1999).

\bibitem{Moore_2016} Takahiro Morimoto, Shudan Zhong, Joseph Orenstein, and Joel E. Moore, Phys. Rev. B \textbf{94}, 245121 (2016).

\bibitem{Duval_2006} C. Duval, Z. Horvth, P. A. Horvthy, L. Martina, and P. C. Stichel, Mod. Phys. Lett. B, \textbf{20}, 373 (2006).

\bibitem{Son_2012} Dam Thanh Son and Naoki Yamamoto, Phys. Rev. Lett., \textbf{109}, 181602 (2012).

\bibitem{Yin_2012} M. A. Stephanov and Y. Yin, Phys. Rev. Lett., \textbf{109}, 162001 (2012).

\bibitem{Chen_2013} Y. Chen, Si Wu, and A. A. Burkov, Phys. Rev. B \textbf{88}, 125105 (2013).

\bibitem{Kenji_2008} Kenji Fukushima, Dmitri E. Kharzeev, and Harmen J. Warringa, Phys. Rev. D \textbf{78}, 074033 (2008).

\bibitem{Kim:2014} Ki-Seok Kim, Heon-Jung Kim, and M. Sasaki, Phys. Rev. B \textbf{89}, 195137, (2014).

\bibitem{Franz_2013} M. M. Vazifeh and M. Franz, Phys. Rev. Lett. \textbf{111}, 027201 (2013).

\bibitem{Basar_2014} G. Basar, D. E. Kharzeev, and H- U. Yee, Phys. Rev. B \textbf{89}, 035142 (2014).

\bibitem{Landsteiner_2014} K. Landsteiner, Phys. Rev. B 89, 075124 (2014).

\bibitem{Pavan_2013} Pavan Hosur, Xiaoliang Qi, Comptes Rendus Physique, \textbf{14}, 857-870 (2013).

\bibitem{Son:2013} D. T. Son and B. Z. Spivak, Phys. Rev. B \textbf{88}, 104412 (2013).

\bibitem{Lundgren:2014} R. Lundgren, P. Laurell, and G. A. Fiete, Phys. Rev. B \textbf{90} 165115 (2014).

\bibitem{Sharma:2016} G. Sharma, P. Goswami, and S. Tewari, Phys. Rev. B \textbf{93}, 035116 (2016).

\end{thebibliography}
\end{document}